# The emergence of magnetic skyrmions

**During the past decade, axisymmetric two-dimensional solitons (so-called *magnetic skyrmions*) have been discovered in several materials. These nanometer-scale chiral objects are being proposed as candidates for novel technological applications, including high-density memory, logic circuits and neuro-inspired computing.**

Alexei N. Bogdanov and Christos Panagopoulos

**R**esearch on magnetic skyrmions benefited from a cascade of breakthroughs in magnetism and nonlinear physics. These magnetic objects are believed to be nanometers-size whirling cylinders, embedded into a magnetically saturated state and magnetized antiparallel along their axis (Fig. 1). They are topologically stable, in the sense that they can't be continuously transformed into homogeneous state. Being highly mobile and the smallest magnetic configurations, skyrmions are promising for applications in the emerging field of spintronics, wherein information is carried by the electron spin further to, or instead of the electron charge. Expectedly, their scientific and technological relevance is fueling inventive research in novel classes of bulk magnetic materials and synthetic architectures [1, 2].

The unconventional and useful features of magnetic skyrmions have become focus of attention in the science and technology circles, sometimes teasingly called "*magic knots*" or "*mysterious particles*". Interestingly, for a long time it remained widely unknown that the key "mystery" surrounding skyrmions is in the very fact of their existence. Indeed, mathematically the majority of physical systems with localized structures similar to skyrmions are radially unstable, collapsing spontaneously into linear singularities. However, in the 1980's a surprising mathematical development indicated magnetic materials with broken mirror symmetry defy this "general rule" of instability [3].

Crucially, in non-centrosymmetric magnets, magnetic interactions imposed by the handedness of the underlying crystallographic structure counteract the collapse and stabilize finite size spin textures [3, 4]. Hence, axisymmetric localized states (*magnetic skyrmions*) form due to a subtle balance between the counteracting internal forces. Such "self-supporting" particle-like objects are known as solitons. By now, it is generally acknowledged that the unique stabilization mechanism and the soliton paradigm govern the physics of magnetic skyrmions. To place the subject into perspective, we shall begin our discussion from a remarkable observation made almost two centuries ago.

**Alexei Bogdanov** is a senior researcher at the Leibniz Institute for Solid State and Materials Research Dresden in Germany and a project professor in the Chirality Research Center at Hiroshima University in Japan. **Christos Panagopoulos** is a professor of physics at Nanyang Technological University and an investigator at the National Research Foundation, both in Singapore.

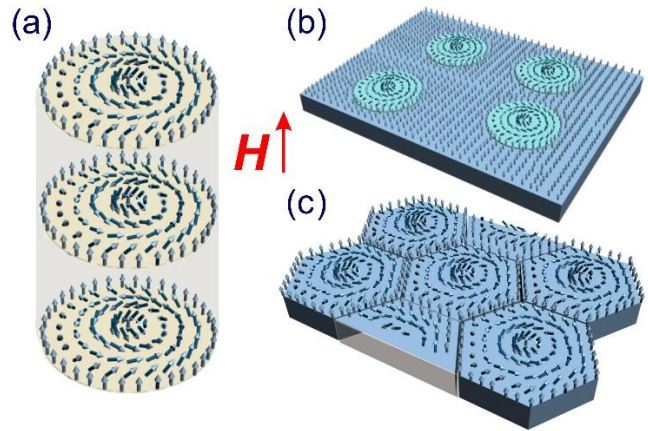

**Figure 1. Magnetic skyrmions** are nanoscale spinning magnetic cylinders **(a)**, embedded into the saturated state of ferromagnets **(b)**. In the skyrmion core, magnetization gradually rotates along radial directions with a fixed rotation sense namely, from antiparallel direction at the axis to a parallel direction at large distances from the center. These are known as *Bloch-type*, one of the *five* possible skyrmion configurations arising in uniaxial ferromagnets [3]. Under certain conditions, isolated skyrmions condense into a hexagonal lattice **(c)**. Panels (a-c) depict skyrmion configuration in cubic helimagnets and uniaxial ferromagnets with $D_n$ symmetry.

## The birth of a *soliton*

In August of 1834, Scottish engineer John Scott Russell realized that the sudden stopping of a boat in a narrow water canal stimulated a well-localized disturbance that propagated virtually unchanged for many miles. He wrote: *"…the mass of water in the channel which it [the boat] had put in motion… rolled forward with great velocity, assuming the form of a large solitary elevation, a rounded, smooth and well-defined heap of water, which continued its course along the channel apparently without change of form or diminution of speed…"* [5].

Inspired by this observation, Russell built wave-banks in his house and continued experiments with such waves of translations (also called *solitary waves*). He was convinced solitary waves are of fundamental importance however, his work was viewed with skepticism, primarily because his findings and conclusions were not in agreement with the properties of water waves described by the theory prevailing at the time. To appreciate the significance and relevance of Russell's discovery, we shall consider the exact solution of the equation describing a solitary wave as an infinite sum (*superposition*) of harmonic waves. It is known that the phase speeds of harmonic water waves depend on their wavelength (*dispersion effect*). Thus, once formed, a travelling solitary wave gradually spreads and eventually decays.

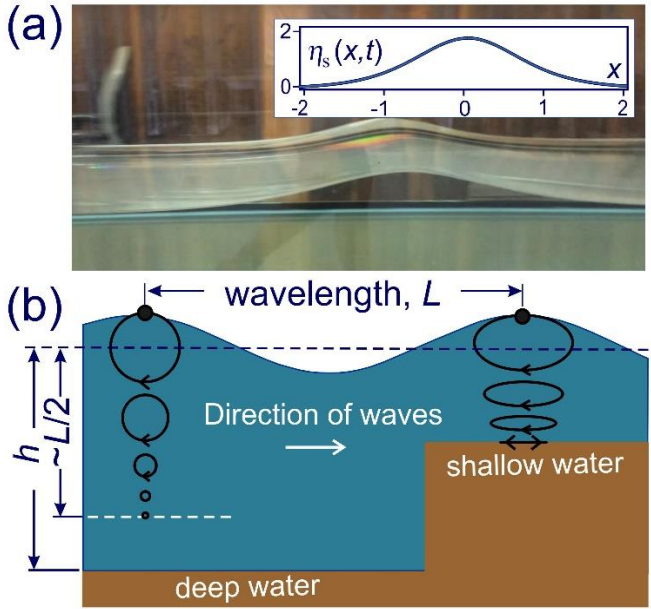

**Figure 2. Russell's solitary-wave solitons** occur in shallow water channels, **(a)** as shown in a laboratory reconstruction of Russell's original observation (Adapted from ref. 6). The inset is a typical localized solution of the Korteweg-de Vries equation for extremely shallow water (Adapted from ref. 5). **(b)** In deep water, waves propagate (from left to right), whereas water particles orbit in circles around their average position. The diameter of the circular orbits shrinks gradually with increasing distance from the surface down to half the wavelength *L*. In shallow waters with depth *h* < *L*/2, the orbits become elliptical and essentially flat at the bottom of the tank or seabed.

The enigmatic stability of solitary waves in Russell's experiments was explained in 1871 by French physicist Joseph Boussinesq, who showed that in shallow waters solitary waves are stabilized by the seabed. Six years later, he introduced an equation to describe this phenomenon [5]:

$$\underbrace{\frac{1}{v_0}\frac{\partial \eta}{\partial t} + \frac{\partial \eta}{\partial x}}_{\text{transport}} + \underbrace{\frac{h^2}{6}\frac{\partial^3 \eta}{\partial x^3}}_{\text{dispersion}} + \underbrace{\frac{3}{2h}\eta\frac{\partial \eta}{\partial x}}_{\substack{\text{interaction}\\\text{with seabed}}} = 0, \quad v_0 = \sqrt{gh}$$

where *t* is time, $\eta$ is function of distance *x* and *t*, *h* is depth, and *g* is gravitational acceleration. It became known as the Korteweg-de Vries (KdV) equation in honor of the physicists who in 1898 published detailed studies of the localized and periodic solutions. The first two terms in the KdV equation, labeled as transport, yield solutions for non-dispersive waves propagating with speed $v_0$. The next term describes the dispersion, and the last term accounts for the interaction with the seabed.

When *h* is a factor of 20 or more smaller than the wavelength *L*, the equation has localized solutions with bell-shaped profiles $\eta_s(x, t)$ [5] shown in Fig. 2(a), that describe solitary waves [6]. Contrary to deep-water waves, the balance of dispersion and the interaction with the seabed keeps shallow solitary waves stable and preserves their bell-shaped profile.

For a long time, contemporary science overlooked this fundamental property of solitary waves, however, they received renewed interest in the 1960's. This was largely inspired by the discovery of localized states in various domains of physics. Since then researchers have used modern mathematical methods and numerical simulations to study solitary waves and other self-supporting localized states—which are referred to as *solitons*, a term coined by Zabusky and Kruskal [7].

## Physics beyond the *linear* world

In the KdV equation, stabilization of solitary waves is due to the quadratic (*nonlinear*) term in variable $\eta$. Other terms are *linear* with respect to $\eta$ and do not support the formation of localized states. *Nonlinearity* is the characteristic feature of soliton mathematics, in contrast to *linear* models applied to most known physical systems [5].

Let us unveil the physical picture behind these mathematical notions. We recall the *superposition principle* (or *superposition property*) namely, the net response caused by two or more stimuli is the sum of responses that would have been caused by each stimulus, separately. Notably, the majority of physical systems comply with the *superposition principle*. A typical example is the weak wave distortions in the surface of water. Hence, when two or more waves overlap in space, the resultant disturbance $\eta(x, t)$ is equal to the sum of the individual disturbances $\eta_i(x, t)$ (i.e. $\eta(x, t) = \eta_1(x, t) + \eta_2(x, t) + \ldots$). In other words, they are expressed as *linear* functions. This means, *linear* models in mathematics correspond to physical systems obeying the superposition principle. Such physical systems are called *linear*.

*Linearity* is ubiquitous in nature and engineering. For example, electromagnetic phenomena are described by the linear Maxwell equations and subatomic physics is described by the linear Schrödinger equation. Furthermore, in physics many weakly distorted systems can be successfully described by *linear* models. However, it is important to understand that such models are in fact approximately linear. In particular, when sufficiently strong distortions violate the superposition principle, the system becomes *nonlinear*, where *the whole is not equal to the sum of its parts*.

To demonstrate how the linear model transforms to nonlinear, we employ the simplified model of the surface wave structure shown in Fig. 2(b). Consider *deep sea water*, where fluid particles under the wave-surface move in circular orbits. The diameter of the orbit decreases with increasing distance from the surface, becoming zero when depth (*h*) is half the wavelength (*L*). There is no wave motion below the wave base (*h* > *L*/2). Now, when *h* < *L*/2 wave motion spreads to the seabed. Here, orbits are initially elliptical, flatten gradually with depth and finally reduce to horizontal oscillations at the seabed. Furthermore, distortions of the particle orbits increase with decreasing depth and at extremely small depths (*h* < *L*/20) become sufficiently strong to suppress the dispersion processes, and stabilize solitary waves. Hence, the stabilization of solitons is supported by strong distortions of the surface wave structures, imposed by the seabed at extremely small depths (*h* < *L*/20) and accounted for by the nonlinear term in the KdV equation.

Therefore, localized solutions $\eta_s(x, t)$ describe travelling solitary waves, which preserve their shape and are robust against perturbations *i.e.* behave as *solid particles*. In other words, the solitary wave solution $\eta_s(x, t)$ introduces novel physical entities namely, *internally stable localized structures with particle-like properties* (*solitons*). (For more on localized states in nonlinear systems, see the article by D. Campbell, S. Flach, and Yu. Kivshar, Physics Today, January 2004, page 43).

The beauty here is that shallow-water solitary waves and the associated mathematical model, represent the extended class of solitons namely, self-supported, particle-like excitations emerging in *nonlinear* physical systems. Next, we introduce magnetic skyrmions and employ this member of the soliton family to link magnetism to the soliton paradigm and non-linearity.

## Stability of magnetic skyrmions

Essentially, magnetic skyrmions are localized axisymmetric configurations of a magnetization vector *M* (Fig. 1). Unlike one-dimensional solitons described by the KdV equation, magnetic skyrmions are two-dimensional solitons where *M* varies (*rotates*) along two spatial directions. Regular solutions for one-dimensional solitons have been derived in many nonlinear equations and investigated with established mathematical methods [6]. This is not the case for two- and three-dimensional solitons. As a matter of fact, it has been mathematically proven that solutions for two- and three-dimensional solitons are unstable in the majority of field-models. Indeed, according to *Derrick-Hobart* theorem, axisymmetric solitons (Fig. 1(a)) in a common ferromagnetic material should shrink and collapse to a linear singularity.

Although this is generally the case, Derrick-Hobart theorem does not apply to field-models where instabilities of multidimensional solitons are eliminated by the *stabilization* functional. The latter is composed of either *(i)* higher order degrees of the spatial derivatives, or *(ii)* terms linear with respect to the first spatial derivatives that is, *Lifshitz invariants* (Box 1). Tony Skyrme, a British mathematical physicist pioneered the investigation of the first-type and in 1961 reported solutions for three-dimensional solitons; low-energy dynamics of mesons and baryons [8]. Solutions for two-dimensional magnetic solitons (aka *vortices* or *skyrmions*) imposed by the Lifshitz invariants were derived in 1989 [3].

A question arising is *whether multi-dimensional solutions derived from the two types of mathematical field-models can describe real physical phenomena*. Well, in condensed matter physics there are no obvious physical interactions described by higher order degrees of spatial derivatives. On the other hand, the second-type can describe magnetic properties of systems with broken inversion symmetry namely, *non-centrosymmetric magnets* (Box 1) [3,9]. In this type of field-models, the stabilization term is composed of *Lifshitz invariants* (Box 1). The good news here, is that this mathematical form reflects magnetic, so-called Dzyaloshinskii-Moriya interactions imposed by the chirality of the crystallographic structure. Importantly, similar interactions apply to a wide range of condensed matter systems possessing broken inversion symmetry. Clearly, this is very practical. To start with, the list of readily available materials includes chiral liquid crystals, multiferroics and metallic nanostructures, all supporting multidimensional solitons.

## The name of the *skyrmion*

It would be useful to comment on terminology especially, considering the polysemantic character of the term "skyrmion" in literature. The term was coined at the beginning of 1980s for a family of multidimensional topological solitons described by the nonlinear field model introduced by Tony Skyrme [8]. Since then, "skyrmion" has been loosely employed to designate a variety of physical phenomena in particle and nuclear physics, string theory, and condensed matter [10]. In condensed matter physics for example, "skyrmion" has been assigned to non-singular localized and topologically stable field configurations, distinguishing them from singular localized states such as disclinations in liquid crystals.

The wide use of the term reminds us of Umberto Eco and his comment why his cult novel was called The Name of the Rose: "*because rose is a symbolic figure, so rich in meanings that by now it hardly has any meaning left*". For skyrmions, such "wealth in meaning" could lead to a fundamental misunderstanding, where these particles might be considered to share common physical properties, or even identified as a fundamental entity (similar to "electrons", "neutrinos", or "quarks"). In its modern usage, "skyrmion" does not describe a specific object. Instead, it is used as *umbrella title*, a label for a broad range of fundamentally different physical phenomena.

## Box 1. Stabilization mechanism

Magnetic skyrmions can emerge in magnetically ordered compounds lacking inversion symmetry (non-centrosymmetric crystals) [3]. Chiral magnets (with broken mirror symmetry) belong to the most common and well-investigated group of non-centrosymmetric ferromagnets hosting magnetic skyrmions. Conveniently, they offer visual access to the physical mechanism underlying magnetic skyrmions. We recall, a chiral object and its mirror image are distinguishable, and create two forms of a crystal: left-handed and right-handed enantiomers. An enantiomorphic pair of the chiral cubic helimagnet FeGe is shown in the figure below.

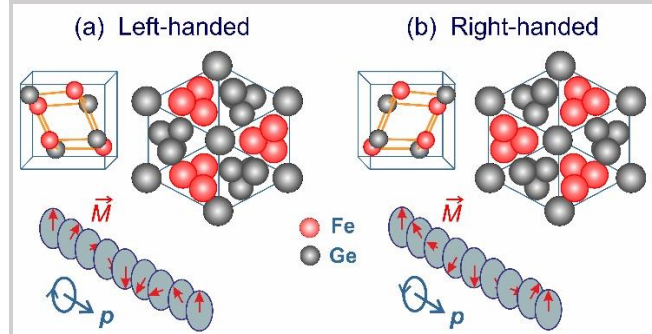

In chiral ferromagnets, the underlying crystallographic structure induces specific magnetic forces (Dzyaloshinskii-Moriya interactions) [4]. In turn, these forces give rise to homochiral spatial modulations of the magnetization rotation in the plane perpendicular to the propagation direction *p* (helices). Such chiral "strings" compose Bloch type magnetic skymions (Fig. 1) [3]. For example, in polar non-centrosymmetric ferromagnets ($C_{nv}$ symmetry), Dzyaloshinskii-Moriya interactions stabilize strings with magnetization rotation along the propagation directions and a fixed sense of rotation (cycloids). These give rise to one of the five possible skyrmion configurations namely, the Neel-type (Fig. 3(a)). The reminder types of magnetic skyrmions are described in Ref. [3] (see also [11]).

Mathematically, Dzyaloshinskii-Moriya interactions are described by energy contributions composed of functionals $\mathcal{L}_{ij}^{(k)} = M_i \partial M_j / \partial x_k - M_j \partial M_i / \partial x_k$ (known as Lifshitz invariants) [4]. Invariants $\mathcal{L}_{ij}^{(k)}$ are linear in the first spatial derivatives of a type favoring spatial modulations with the magnetization rotation in the *ij*-planes propagating along the *k*- axis. For instance, $\mathcal{L}_{xy}^{(z)}$-invariant stabilizes helices with magnetization rotation in the *xy*-plane, propagating along the *z*-axis. Consider a cubic chiral ferromagnet. Here, the Dzyaloshinskii-Moriya energy

$$w_D(\boldsymbol{M}) = D\left(\mathcal{L}_{yx}^{(z)} + \mathcal{L}_{xz}^{(y)} + \mathcal{L}_{zy}^{(x)}\right) = D(\boldsymbol{M} \cdot \nabla \times \boldsymbol{M}) \quad (B1)$$

favors homochiral helices propagating along three spatial axes with a rotation direction determined by the sign of constant *D* ($\nabla = (\partial/\partial x) e_x + (\partial/\partial y) e_y + (\partial/\partial z) e_z$ is a vector differential operator).

The physical phenomenon known as a "magnetic skyrmion" has been introduced under the terms "magnetic vortex" or "two-dimensional topological soliton" [3]. They can be defined as stable strings, which are (*i*) localized, (*ii*) axisymmetric, with (*iii*) a fixed sense of rotation. They enjoy nontrivial topology, protecting them from unwinding into homogeneous states. However, contrary to common belief, topology does not prevent them from collapsing. We recall, in centrosymmetric ferromagnets topological protection exists only in the form of linear singularities. In a non-centrosymmetric system, the existence of skyrmions with finite sizes is enabled by specific stabilization mechanism. Importantly, for magnetic skyrmions in condensed matter physics the notion of "soliton" is key to elucidating their nature namely, internally stable particle-like objects.

## Isolated skyrmions and lattices

Experimentalists discovered Bloch-type skyrmions in many chiral cubic helimagnets, including FeGe, manganese silicide, MnSi, and copper-selenium oxide, $Cu_2OSeO_3$ [1, 2, 13]. Neel-type skyrmions have appeared in polar non-centrosymmetric ferromagnet gallium-vanadium sulphide, $GaV_4S_8$, and in nanolayers with engineered Dzyaloshinskii-Moriya interactions [1, 12, 13, 16].

A simplified model for the energy $w_0$ of a chiral cubic ferromagnet in an applied magnetic field $H$

$$w_0(M) = \underbrace{A(\nabla \cdot M)^2}_{\text{exchange}} - M \cdot H - \underbrace{D(M \cdot \nabla \times M)}_{\substack{\text{Dzyaloshinskii–Moriya} \\ \text{interaction}}}$$

includes main energy contributions, which play key role in the formation of magnetic skyrmions. The first ferromagnetic exchange energy term, with constant $A$, imposes parallel ordering of the magnetic moments. The second term, which is the interaction with the applied magnetic field, favors magnetization oriented along $H$. The final Dzyaloshinskii-Moriya term induces helical modulations. The model introduces two fundamental parameters: the helix period, $L_D = 4\pi A/|D|$ and the saturation field $H_D = D^2M/(2A)$ suppressing chiral modulations [1,9]. For the two most studied cubic helimagnets, MnSi and FeGe, $L_D$ = 18 and 70 nm and $\mu_0 H_D$ = 620 and 359 mT, in terms of the vacuum permeability $\mu_0$.

At high magnetic fields, minimizing the energy functional $w(M)$ yields isolated skyrmions as weakly repulsive localized states in an otherwise uniformly magnetized state (Fig. 1(b)). They arise from a subtle balance of the competing magnetic forces. Similar to solitary waves and other solitons, isolated magnetic skyrmions manifest as internally stable, particle-like objects in a continuous medium. Numerical simulations demonstrate billiards with colliding and scattering magnetic skyrmions in narrow channels [14].

Figure 4(b) depicts the skyrmion core diameter $L_S$ as a function of applied magnetic field. The antiparallel magnetization at the skyrmion center is energetically unfavourable in the presence of an applied magnetic field, so the skyrmion core gradually shrinks with increasing magnetic field. Below a transition field $H_S$ = 0.801 $H_D$, isolated skyrmions condense into a lattice, although the scenario is realized only if the isolated skyrmions can nucleate. Otherwise, isolated skyrmions persist below $H_S$, elongate, and expand into a one-dimensional band below the elliptic instability field, $H_{el}$ = 0.534 $H_D$. Experimentally, skyrmions do gradually elongate from circular to elliptic to one-dimensional modulations in FePd/Ir nanolayers in a decreasing magnetic field [12, 13], as shown in spin-polarized scanning tunneling microscopy images (Fig. 3(c-g)).

In model $w_0$, isolated skyrmions exist only in applied magnetic fields larger than the elliptic instability field. But in non-centrosymmetric ferromagnetic bulk crystals and artificially synthesized magnetic nanolayers and multilayers, internal magnetic interactions control the preferential direction of the magnetization orientation—called magnetic anisotropy—which influences the area that can host isolated magnetic skyrmions. The interactions can even stabilize skyrmions in the absence of an applied magnetic field [3, 13], as has been observed in FePt/Ir nanolayers with strong uniaxial magnetic anisotropy [16].

With decreasing magnetic field, isolated skyrmions often transform into a lattice, usually with hexagonal symmetry. A demonstration on FeGe nanolayers, as shown in Fig. 4, reveals transitions between a skyrmion lattice and competing one-dimensional phases [9,12, 14].

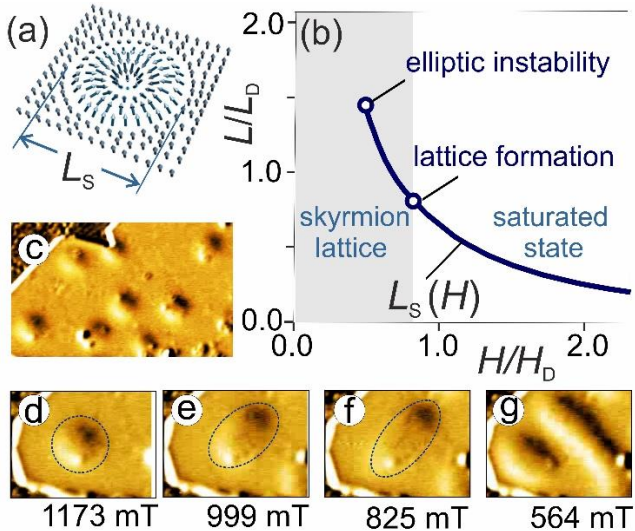

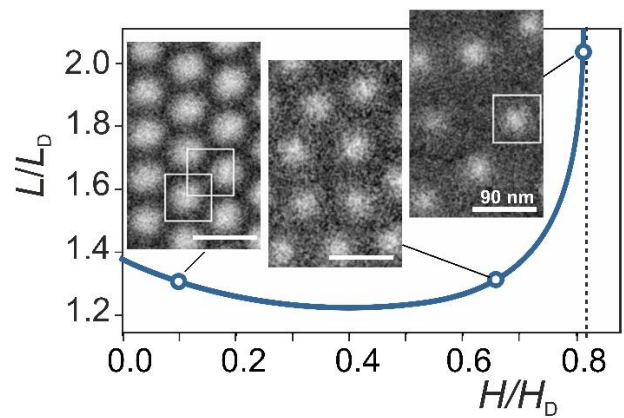

**Figure 3. Isolated Neel-type skyrmions** have magnetization rotating along the direction of propagation (a). **(b)** The calculated skyrmion diameter $L_S$ shrinks as the applied magnetic field increases. The characteristic parameters $L_D$ and $H_D$ are the helix period and saturation field, respectively. **(c)** Skyrmions in an iron-palladium-iridium nanolayer are isolated at an applied magnetic field $\mu_0 H$ = 1280 mT. With decreasing magnetic field, a skyrmion (circled) deforms into an ellipse and terminates as a one-dimensional modulation **(d–g)**. (Adapted from ref. 11.)

**Figure 4. A hexagonal skyrmion lattice** evolves in the presence of weaker applied magnetic fields in materials that allow nucleation. The calculated size $L$ of the skyrmion cell varies with the reduced applied magnetic field, $H/H_D$. The characteristic parameters $L_D$ and $H_D$ are the helix period and saturation field, respectively. The insets are three electron holography images of skyrmion lattices in a thin layer of iron monogermanide for different applied magnetic fields—from left to right, 100 mT, 350 mT, and 400 mT. (Adapted from ref. 15.)

The flexibility to tune spin textures invited intensive experimental and theoretical investigations in non-centrosymmetric ferromagnets and magnetic nanolayers and multilayers. A broad spectrum of unique properties of magnetic skyrmions has already emerged - for example, their creation, deletion, and motion with a magnetic tip [12].

## Perspective and potential technology

Magnetic skyrmions constitute a promising new direction for data storage and spintronics. They are countable objects that can be created and manipulated in, for example, layers of magnetically soft materials to develop versatile nanoscale magnetic patterns [2, 16]. Skyrmions thus offer a route to localized magnetic inhomogeneities in low-coercivity materials, which are not viable for traditional magnetic recording.

Interactions similar to Dzyaloshinskii–Moriya arise in a wide range of condensed-matter systems with broken inversion symmetry. The introduction of chiral interactions in those systems provides the stabilization mechanism for solitonic states analogous to magnetic skyrmions. For example, axisymmetric solitons appear in chiral liquid crystals [17], a state of matter thermodynamically located between an isotropic liquid and a three-dimensional ordered solid. In ferroelectrics, which display spontaneous electric polarization, layered oxide architectures host emerging magnetic solitons [18]. Non-centrosymmetric condensed-matter systems broadly form a class of materials with multidimensional solitonic states [9].

For condensed-matter physicists, Dzyaloshinskii–Moriya interaction offers a playground for investigations that only require lack of inversion symmetry. In addition to naturally occurring systems, such as chiral cubic helimagnets, researchers can engineer a structure that cannot be inverted—for example, the interface between two materials. The interface between a ferromagnet and a strong spin-orbit metal gives rise to tunable-strength Dzyaloshinskii–Moriya interactions, which vary with the material selection [1, 2, 16]. The interfacial effects can even dominate if the ferromagnet is sufficiently thin.

The flexibility to design the host and tune the skyrmion properties offers versatility for technological applications. Skyrmion-based devices have the potential to store and process information in unprecedentedly small spaces [1, 2, 14, 16]. The presence or absence of a skyrmion could serve as a 1 or 0 in a data bit for racetrack memory, and multiple skyrmions could aggregate to form storage devices. The states of such devices could be modulated by an electric current that drives skyrmions in and out of the devices, analogous to biological synapses. The devices could thus potentially perform neuromorphic pattern-recognition computing.

Researchers have already engineered interfacial skyrmions in magnetic multilayers at up to room temperature [16]. Those skyrmions offer an opportunity to bring topology into consumer-friendly nanoscale electronics. In the magnetic multilayers, exotic skyrmion configurations also form under the combined influence of chiral interactions and magnetodipolar effects. Such skyrmion hybrids present a novel class of localized states that have yet to be explored in depth.

But there are still fundamental properties to investigate, such as the wave–particle duality of skyrmions, their interaction with other magnetic textures, and skyrmion lattices as magnonic crystals. The interaction between magnetic skyrmions and light or other topological excitations—for example, superconducting vortices—could also lead to new exotic states of matter. Investigating those topics will require new material architectures and advancements in characterization techniques.

*A.N.B. thanks M. Ochi and K. Inoue for hospitality and collaboration during his stay at Hiroshima University. C.P. thanks the members of his research group for insightful discussions. This work is supported in Germany by the Deutscher Forschungsgemeinschaft through SPP2137 "Skyrmionics", and in Singapore by Academic Research Fund Tier 3 (Reference No. MOE5093) and the National Research Foundation (Reference No. NRF-NRFI2015-04).*